\documentclass[11pt,a4paper]{article}
\usepackage{jcappub}
\usepackage{graphicx}
\usepackage{color}
\usepackage{latexsym}
\usepackage{psfrag}
\usepackage{epsfig}
\usepackage{overpic}

\newcommand{\be}{\begin{eqnarray}}
\newcommand{\ee}{\end{eqnarray}}
\newcommand{\rar}{\rightarrow}

\title{Testing the Kerr metric with the iron line and the KRZ parametrization}

\author[a]{Yueying~Ni,}
\author[a]{Jiachen~Jiang,}
\author[a,b,1]{Cosimo~Bambi,%
\note{Corresponding author}}

\affiliation[a]{Center for Field Theory and Particle Physics and Department of Physics,\\
Fudan University, 220 Handan Road, 200433 Shanghai, China}
\affiliation[b]{Theoretical Astrophysics, Eberhard-Karls Universit\"at T\"ubingen,\\ 
Auf der Morgenstelle 10, 72076 T\"ubingen, Germany}

\emailAdd{yyni13@fudan.edu.cn}
\emailAdd{jcjiang12@fudan.edu.cn}
\emailAdd{bambi@fudan.edu.cn}

\abstract{The spacetime geometry around astrophysical black holes is supposed to be well approximated by the Kerr metric, but deviations from the Kerr solution are predicted in a number of scenarios involving new physics. Broad iron K$\alpha$ lines are commonly observed in the X-ray spectrum of black holes and originate by X-ray fluorescence of the inner accretion disk. The profile of the iron line is sensitively affected by the spacetime geometry in the strong gravity region and can be used to test the Kerr black hole hypothesis. In this paper, we extend previous work in the literature. In particular: $i)$ as test-metric, we employ the parametrization recently proposed by Konoplya, Rezzolla, and Zhidenko, which has a number of subtle advantages with respect to the existing approaches; $ii)$ we perform simulations with specific X-ray missions, and we consider NuSTAR as a prototype of current observational facilities and eXTP as an example of the next generation of X-ray observatories. We find a significant difference between the constraining power of NuSTAR and eXTP. With NuSTAR, it is difficult or impossible to constrain deviations from the Kerr metric. With eXTP, in most cases we can obtain quite stringent constraints (modulo we have the correct astrophysical model).}


\begin{document}

\maketitle

%%%%%%%%%%%%%%%%%%%%%%%%%%%%%%%

\section{Introduction}

In 4-dimensional general relativity, uncharged black holes are described by the Kerr solution and are completely specified by only two parameters, associated with the mass $M$ and the spin angular momentum $J$ of the compact object~\cite{h1,h2,h3}. There are some assumptions behind this statement. The spacetime must be stationary and asymptotically flat. The exterior is empty (vacuum) and regular (no singularities or closed time-like curves). For a review, see e.g. Ref.~\cite{h4}.

It is remarkable that the spacetime around astrophysical black holes should be well approximated by the Kerr solution. Initial deviations from the Kerr metric are quickly radiated away with the emission of gravitational waves~\cite{h5}. The equilibrium electric charge is reached very quickly because of the highly ionized host environment of this objects and its impact is completely negligible in the spacetime metric~\cite{h6}. Accretion disks have a low density and their mass is typically many orders of magnitude smaller than the black hole, so their presence cannot appreciably change the background metric~\cite{h7,h8}. In the end, deviations from the Kerr metric can only be expected in the presence of new physics.

Tests of the Kerr metric are motivated by a number of theoretical models. Non-Kerr black hole solutions typically show up in extensions of general relativity~\cite{ex1,ex2,ex3,ex4}. ``Hairy'' black holes are possible in Einstein's gravity in the presence of exotic matter~\cite{hbh1,hbh2,hbh3,hbh4}. Macroscopic deviations from the Kerr metric may also be generated by quantum gravity effects~\cite{q1,q2,q3}

In the past few years, there have been significant efforts to study how to test the Kerr black hole hypothesis with electromagnetic radiation~\cite{r1,r2} and gravitational waves~\cite{r3}. In the case of the electromagnetic approach, the two leading techniques to probe the nature of astrophysical black holes are the analysis of the thermal spectrum (continuum-fitting method)~\cite{cfm1,cfm2} and the analysis of the reflection spectrum (iron line method)~\cite{i1,i2}. Both the techniques were originally proposed, and later developed, to measure black hole spins under the assumption of the Kerr metric and more recently have been studied to test the Kerr black hole hypothesis~\cite{k1,k2,k3,k4,k5,k6,k7,k8,k9,k10,k11,k12,k13,k14,k15}. In the presence of high quality data and with the correct astrophysical model, the iron line method is a more power tool than the continuum-fitting method and can potentially provide superb constraints on possible deviations from the Kerr solution~\cite{ii1,ii2,ii3,ii4}.

Current efforts aiming at inferring model-independent constraints on possible deviations from the Kerr metric with electromagnetic radiation follow the spirit of the parametrized Post-Newtonian (PPN) approach~\cite{ppn}, commonly employed in Solar System tests. In the PPN case, one wants to test the Schwarzschild solution in the weak field limit and adopts the most general static and spherically symmetric line element based on an expansion in $M/r$. The value of the coefficients in front of the expansion parameter has to be measured by observations, and one can check {\it a posteriori} whether they match with those expected for the Schwarzschild metric. In the case of tests of the Kerr metric in the strong gravity region, there are a number of complications, because it is not possible to perform an expansion in $M/r$ and there is currently no satisfactory formalism to test astrophysical black hole in a model independent way. There are several proposals in the literature~\cite{p1,p2,p3,p4,p5,p6,p7}, each of them with its advantages and disadvantages, and the search for a more suitable metric is still a work in progress.

In this paper, we continue our study to use the iron line method to test the Kerr black hole hypothesis. As test-metric, we employ for the first time the Konoplya-Rezzolla-Zhidenko (KRZ) parametrization, which has been recently proposed in Ref.~\cite{krz} and has a number of subtle advantages with respect to the other test-metrics currently considered in the literature. We simulate observations with NuSTAR\footnote{http://www.nustar.caltech.edu}, as an example of a current X-ray mission, and eXTP\footnote{http://www.isdc.unige.ch/extp/}, to illustrate the constraining power of the next generation of observational facilities. We treat the simulated observations as real data and we use XSPEC\footnote{https://heasarc.gsfc.nasa.gov/xanadu/xspec/} to analyze the iron K$\alpha$ line and constrain the KRZ deformation parameters.

Our simulations clearly show a significant difference between NuSTAR and eXTP. We find that it is difficult or impossible to constrain the KRZ deformation parameters with NuSTAR. Large deviations from the Kerr solutions cannot be ruled out, and this is basically true for all the parameters. On the contrary, the constraints inferred from the simulated observations with eXTP are surprisingly strong for all the KRZ deformation parameters except one ($\delta_6$). The difference in the results is due to the unprecedented large effective area of the LAD instrument on board of eXTP. However, such precise measurements of the deformation parameters inevitably require sophisticated astrophysical models to properly describe the reflection spectrum. At the moment there is not a common consensus on the actual possibility of having all systematics under control to reach such a level of precision in the measurements.

The paper is organized as follows. In Section~\ref{s-krz}, we review the KRZ parametrization. In Section~\ref{s-iron}, we briefly discuss the physics of the broad iron lines observed in the reflection spectrum of astrophysical black holes, we provide a short description of our code, and we compute the shape of the iron line in the presence of non-vanishing KRZ deformation parameters. In Section~\ref{s-sim}, we simulate observations with NuSTAR and eXTP, and we analyze the data with XSPEC to constrain the KRZ deformation parameters. Section~\ref{s-last} provides a summary of this work and remarks our main results. Throughout the paper, we employ natural units in which $G_{\rm N} = c = 1$ and the convention of a metric with signature $(- + + +)$.

\section{Konoplya-Rezzolla-Zhidenko parametrization \label{s-krz}}

In order to test the nature of astrophysical black holes, we could consider some alternative theory of gravity, find its black hole solutions, and then check whether observations prefer the hypothesis that a certain object is a Kerr black hole of general relativity or a black hole of the alternative theory of gravity. Such an approach meets two problems. First, it is usually difficult to obtain exact rotating black hole solutions in alternative theories of gravity: the spin plays an important rule and it is not very useful to have the non-rotating solution. Second, there are many alternative theories of gravity, no theory seems to be much more motivated than the others, and we may also expect that they are all toy models. If general relativity is not the correct theory to describe black holes, it is possible that we do not have the correct theory at the moment. These considerations have led many authors to employ parametrized metrics. With such approach, one considers a metric in which deviations from the Kerr solutions are quantified by a number of parameters and tries to measure the values of these parameters to check whether observations require that their value is the one expected in the Kerr metric.

Even the parametrized approach meets some difficulties. There are now several proposal in the literature~\cite{p1,p2,p3,p4,p5,p6,p7}, each of them with its advantages and disadvantages. Recently, Konoplya, Rezzolla, and Zhidenko have proposed a new parametrization in Ref.~\cite{krz}. Such a proposal tries to address the following issues (see Ref.~\cite{krz} for more details):
\begin{enumerate}
\item Many parametrizations are based on an expansion in $M/r$. If we want to test the strong gravity region near the black hole horizon, where $M/r$ is not a small quantity, we have an infinite number of roughly equally important parameters, which makes it impossible to isolate the dominant terms and focus the efforts on the measurement of a small number of parameters.
\item One would like to have a so general parametrization that it is possible to recover any black hole solution in any (known and unknown) alternative theories of gravity for specific values of its free parameters. While it is difficult to assert how general a parametrization can be, several proposals in the literature fail to recover the known black hole solutions in alternative theories of gravity.
\item Astrophysical black holes may have a non-negligible spin angular momentum, which plays an important rule in the features of the electromagnetic spectrum associated to the strong gravity region and therefore the spin cannot be ignored in any test of the Kerr metric. The Newman-Janis algorithm is a simple trick to obtain a rotating black hole solution from the non-rotating one. However, it is not guaranteed that such an algorithm works beyond general relativity. In the literature there are some examples in which the algorithm works~\cite{zxc1,zxc2,zxc3}, as well examples in which it fails~\cite{yunes}. It is thus unclear whether a parametrization based on the Newman-Janis algorithm can be employed to test astrophysical black holes.   
\end{enumerate}

The KRZ metric addresses the three issues above in the following way. First, it is not based on an expansion in $M/r$. There is a hierarchical structure in the deviations from the Kerr spacetime, so that higher order terms necessarily provide smaller and smaller corrections. Second, the authors explicitly show that their proposal can well approximate with a few parameters the metrics of rotating dilaton black holes and of rotating black holes in Einstein-dilaton-Gauss-Bonner gravity. For example, this is not possible with the very successful metric proposed in~\cite{p3}, which has been extensively studied in the literature. Third, the KRZ metric is not obtained from the Newman-Janis algorithm, because the initial ansatz is already suitable to describe rotating black holes. While the KRZ metric seems to be at first more complicated than previous proposals, it has some nice properties and is worth studying its applicability to test astrophysical black holes from observations of their electromagnetic spectrum.

Assuming reflection symmetry across the equatorial plane and neglecting coefficients of higher orders, the line element of the KRZ metric reads~\cite{krz}
\be
ds^2 &=& - \frac{N^2 - W^2 \sin^2\theta}{K^2} \, dt^2 - 2 W r \sin^2\theta \, dt \, d\phi
+ K^2 r^2 \sin^2\theta \, d\phi^2 
\nonumber\\ &&
+ \frac{\Sigma \, B^2}{N^2} \, dr^2 + \Sigma \, r^2 \, d\theta^2 \, ,
\ee
where
\be
N^2 &=& \left(1 - \frac{r_0}{r}\right) \left[1 - \frac{\epsilon_0 r_0}{r} 
+ \left(k_{00} - \epsilon_0\right)\frac{r_0^2}{r^2} + \frac{\delta_1 r^3_0}{r^3}\right] 
+ \left[ \left(k_{21} + a_{20} \right) \frac{r^3_0}{r^3} 
+ \frac{a_{21} r^4_0}{r^4} \right] \cos^2\theta \, , \nonumber\\
B &=& 1 + \frac{\delta_4 r^2_0}{r^2} + \frac{\delta_5 r^2_0}{r^2} \cos^2\theta \, , \nonumber\\
\Sigma &=& 1 + \frac{a_*^2}{r^2} \cos^2\theta \, , \nonumber\\
W &=& \frac{1}{\Sigma} \left[\frac{w_{00} r^2_0}{r^2} + \frac{\delta_2 r^3_0}{r^3}
+ \frac{\delta_3 r^3_0}{r^3} \cos^2\theta \right] \, , \nonumber\\
K^2 &=& 1 + \frac{a_* W}{r} + \frac{1}{\Sigma} \left( \frac{k_{00} r^2_0}{r^2} 
+ \frac{k_{21} r^3_0}{r^3} \cos^2\theta \right)\, .
\ee
For our purpose, it is convenient to introduce six {\it deformation parameters} $\{ \delta_j \}$ ($j = 1, 2, ... 6$), which are related to the coefficient $r_0$, $a_{20}$, $a_{21}$, $\epsilon_0$, $k_{00}$, $k_{21}$, and $w_{00}$ appearing in the KRZ metric by the following relations
\be
&&r_0 = 1 + \sqrt{1 - a_*^2} \, , \qquad
a_{20} = \frac{2 a_*^2}{r^3_0} \, , \qquad
a_{21} = - \frac{a_*^4}{r^4_0} + \delta_6 \, , \qquad
\epsilon_0 = \frac{2 - r_0}{r_0} \, , \nonumber\\
&&k_{00} = \frac{a_*^2}{r_0^2} \, , \qquad
k_{21} = \frac{a_*^4}{r_0^4} - \frac{2 a_*^2}{r^3_0} - \delta_6 \, , \qquad
w_{00} = \frac{2 a_*}{r^2_0} \, .
\ee
Here the mass is $M=1$ and $a_*$ is the spin parameter. $r_0$ is the radial coordinate in the equatorial plane of the event horizon. The physical interpretation of the deformation parameters can be summarized as follows (see Ref.~\cite{krz} for more details):
\be
& \delta_1 & \rar \;\;\; \text{related to deformations of $g_{tt}$,}
\nonumber\\ 
& \delta_2 , \, \delta_3 & \rar \;\;\; \text{related to rotational deformations of the metric,}
\nonumber\\ 
& \delta_4 , \, \delta_5 & \rar \;\;\; \text{related to deformations of $g_{rr}$,}
\nonumber\\
& \delta_6 & \rar \;\;\; \text{related to deformations of the event horizon.}
\nonumber
\ee
With our choice, the mass-quadrupole moment is the same as in the Kerr metric, and deviations from the Kerr solution are only possible in the strong gravity region.

\section{X-ray reflection spectroscopy \label{s-iron}}

Within the disk-corona model~\cite{corona1,corona2}, a black hole is surrounded by a geometrically thin and optically thick accretion disk, which radiates as a blackbody locally and a multi-color blackbody when integrated radially. The corona is a hot ($k_{\rm B}T \sim 100$~keV), usually optically thin, electron cloud enshrouding the disk. Due to the inverse Compton scattering of the thermal photons from the disk off the high energy electrons in the corona, the latter acts as an X-ray source and has a power-law spectrum. A fraction of these X-ray photons illuminates the disk, producing a reflection component with some fluorescence emission lines, the most prominent of which is usually the iron K$\alpha$ line~\cite{refl}.

The iron K$\alpha$ line is intrinsically narrow. It is peaked at 6.4~keV in the case of neutral iron and shifts up to 6.97~keV in the case of H-like iron ions. The line observed in the reflection spectrum of black holes can instead be very broad, as a consequence of special and general relativistic effects (Doppler boosting, gravitational redshift, light bending) occurring in the strong gravity region around the compact object. In the present of high quality data and the correct astrophysical model, the analysis of the iron K$\alpha$ line can be used to measure the black hole spin (if we assume the Kerr metric) or even probe the spacetime geometry around the compact object. Actually one has to fit the whole reflection spectrum, not only the iron line, but most of the information on the spacetime metric is encoded in the iron line. For this reason the technique is often called iron line method and in this explorative work we restrict our attention to the iron K$\alpha$ line only.

For the calculation of the iron line, we use the code described in Refs.~\cite{k6,k7} with the KRZ metric reviewed in the previous section. Here we briefly summarize the calculation procedure of our code, and we refer the reader to Refs.~\cite{k6,k7} for more details. The photon flux in the flat region (in, e.g., units of photons/s/cm$^2$/keV) can be written as
\be\label{eq-nph}
N_{E_{\rm obs}} = \frac{1}{E_{\rm obs}} \int I_{\rm obs} (E_{\rm obs}) d\tilde{\Omega} 
= \frac{1}{E_{\rm obs}} \int g^3 I_{\rm e} (E_{\rm e}) d\tilde{\Omega} \, ,
\ee
where $I_{\rm obs}$ and $E_{\rm obs}$ are, respectively, the specific intensity of the radiation and the photon energy at infinity. $d\tilde{\Omega}$ is the infinitesimal solid angle subtended by the image of the disk in the observer's sky. $I_{\rm e}$ and $E_{\rm e}$ are, respectively, the specific intensity of the radiation and the photon energy in the rest frame of the gas. $g = E_{\rm obs}/E_{\rm e}$ is the redshift factor and $I_{\rm obs} = g^3 I_{\rm e}$ follows from Liouville's theorem.

The image plane of the distant observer is divided into a number of small elements. For each element, we consider a photon with momentum perpendicular to the image plane and we compute backwards in time the photon trajectory from the point of detection in the image plane to the point of emission in the disk. In this way, we reconstruct the apparent image of the disk in the observer's sky. Each point of the apparent image of the disk is characterized by its redshift factor $g$, which depends on the point of emission in the disk and the photon constants of motion 
\be
g = \frac{\sqrt{ -g_{tt} - 2 \Omega g_{t\phi} - \Omega^2 g_{\phi\phi}}}{1 + \lambda \Omega} \, .
\ee
Here $\Omega$ is the Keplerian angular momentum of the gas in the disk and $\lambda = k_\phi/k_t$ is a constant of motion along the photon trajectory, where $k_\phi$ and $k_t$ are, respectively, the $\phi$ and $t$ components of the photon 4-momentum. After the calculation of all the photon trajectories, we perform the integral in Eq.~(\ref{eq-nph}) and we obtain the spectrum of the line.

For the disk, we employ the Novikov-Thorne model~\cite{ntm}, which is the standard framework for the description of geometrically thin and optically thick disks. The disk is in the equatorial plane perpendicular to the black hole spin. The particles of the gas follow nearly geodesic circular orbits. The inner edge of the disk is at the radius of the innermost stable circular orbit, $r_{\rm in} = r_{\rm ISCO}$. The outer radius is expected at some very large radius, where the emission of the reflection component becomes negligible. In the present paper, we set $r_{\rm out} = r_{\rm ISCO} + 250$~$M$, which is large enough that its exact value is not very important. For the validity of the Novikov-Thorne model, see e.g.~\cite{r2} and reference therein.

The shape of the iron line does depend on the background metric (spin parameter $a_*$ and possible non-vanishing deformation parameters), the inclination angle of the disk with respect to the line of sight of the observer ($i$), and the intensity profile. For the sake of simplicity, here we assume $I_{\rm e} \propto 1/r^3_{\rm e}$, where $r_{\rm e}$ is the radial coordinate of the emission point. Such a choice of the intensity profile corresponds to that expected at large radii in the Newtonian limit (no light bending) within the lamppost corona geometry.

\begin{figure}[t]
\vspace{0.4cm}
\begin{center}
\includegraphics[scale=0.6]{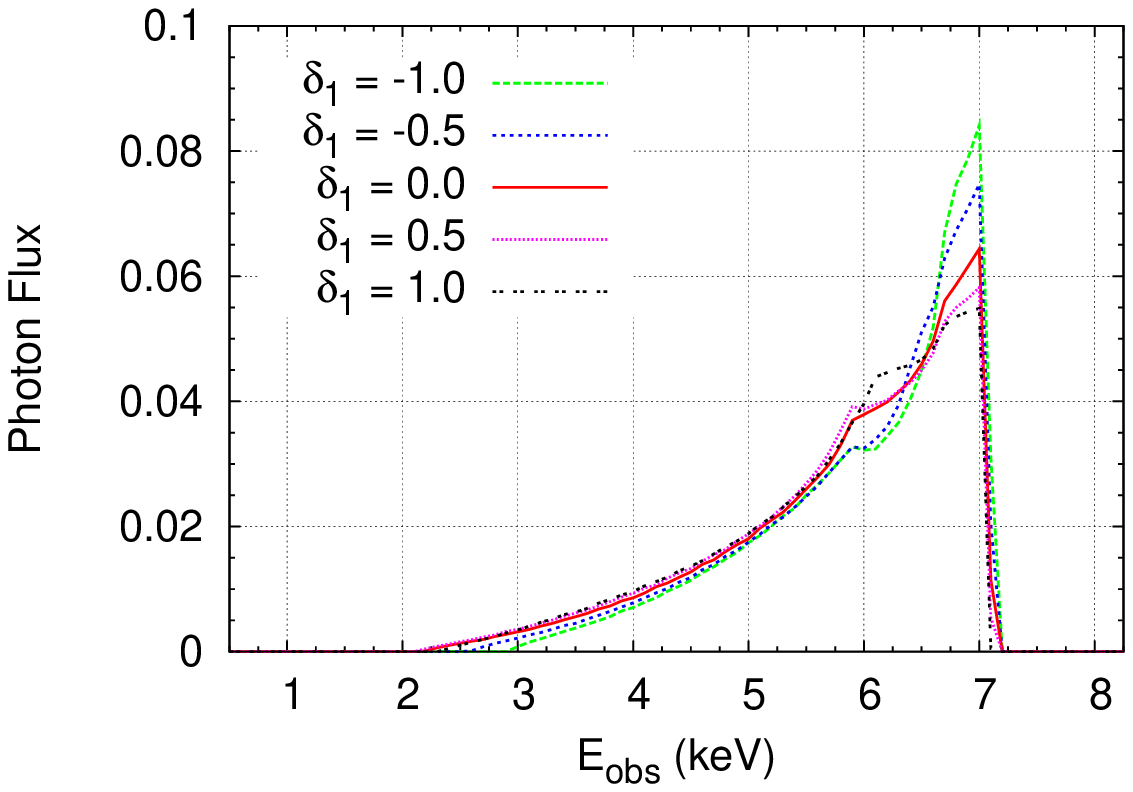}
\includegraphics[scale=0.6]{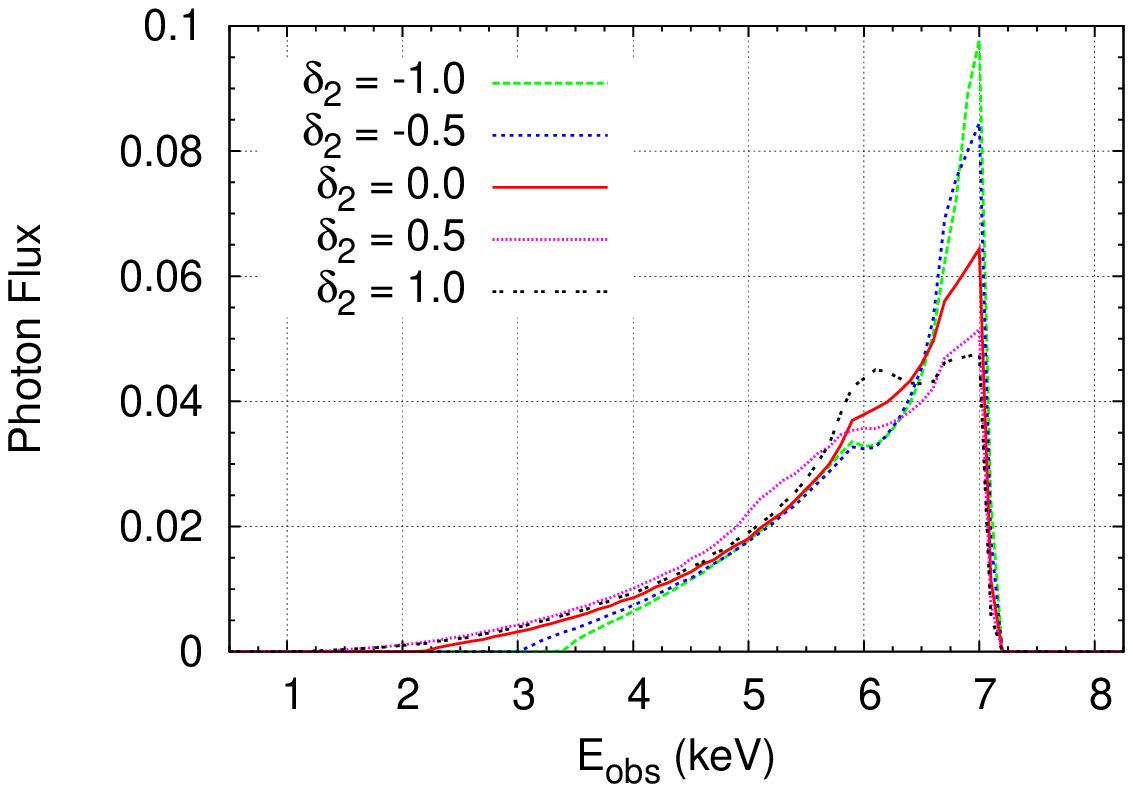} \\
\includegraphics[scale=0.6]{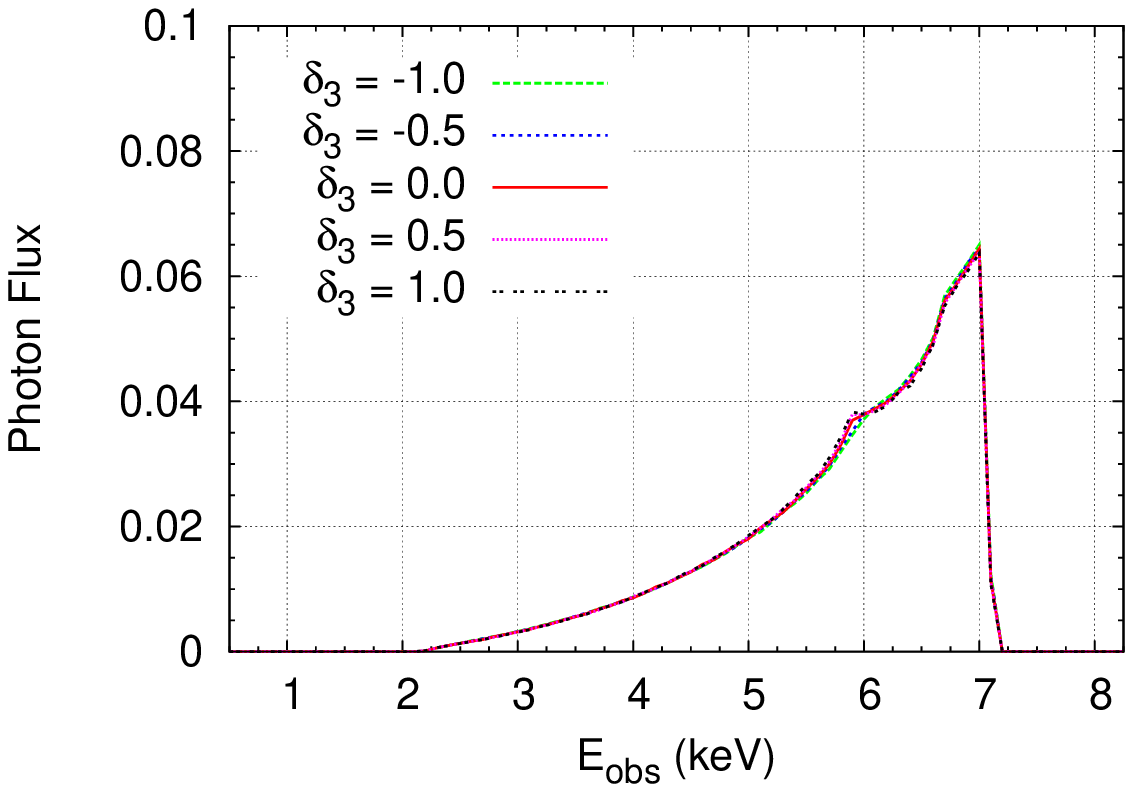}
\includegraphics[scale=0.6]{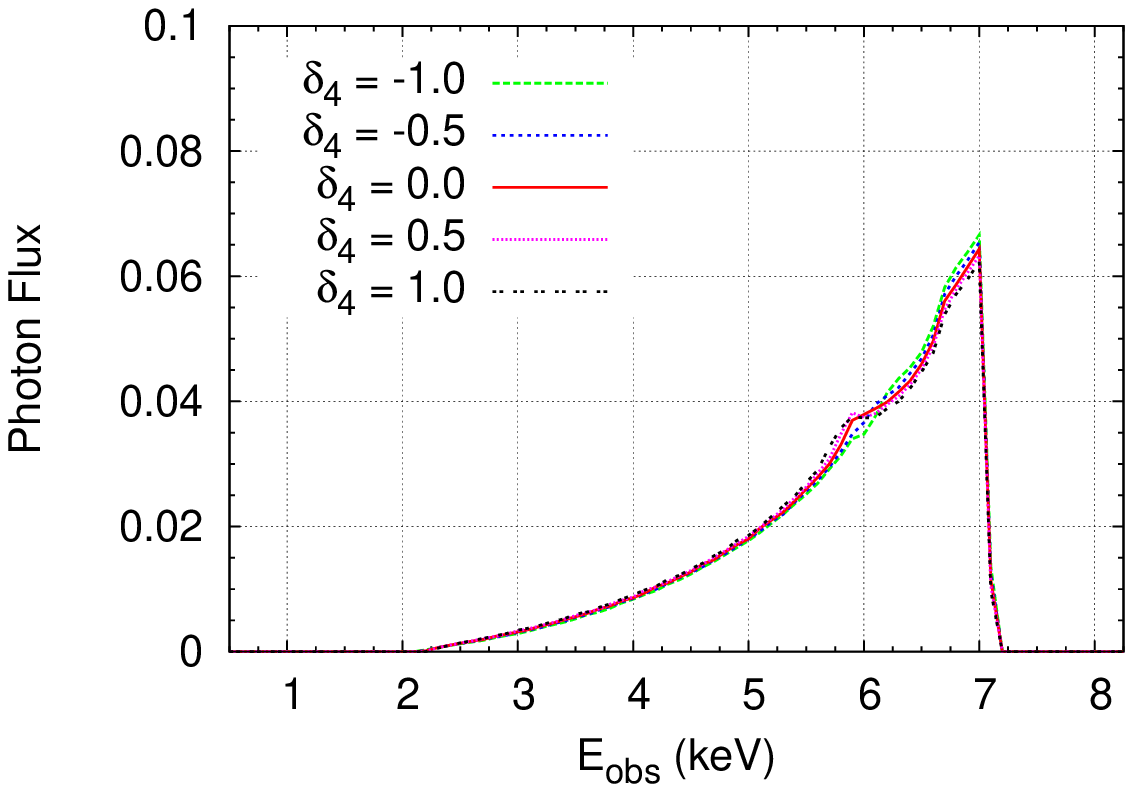} \\
\includegraphics[scale=0.6]{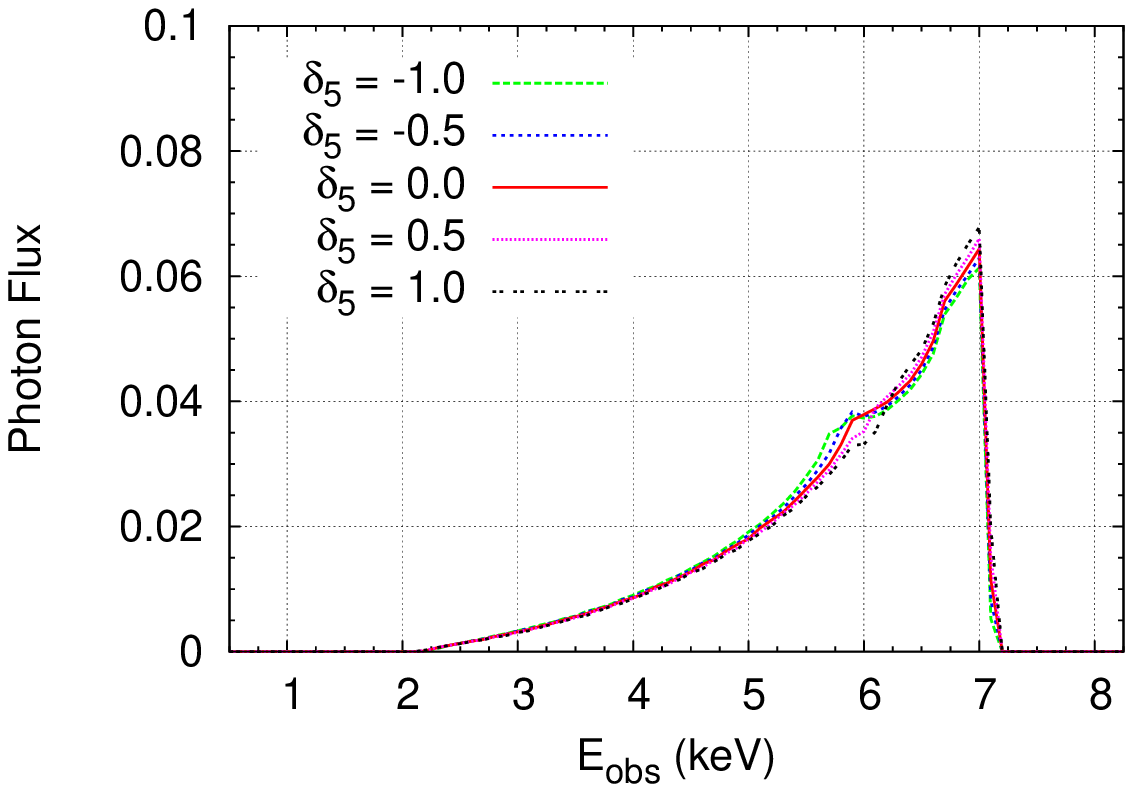}
\includegraphics[scale=0.6]{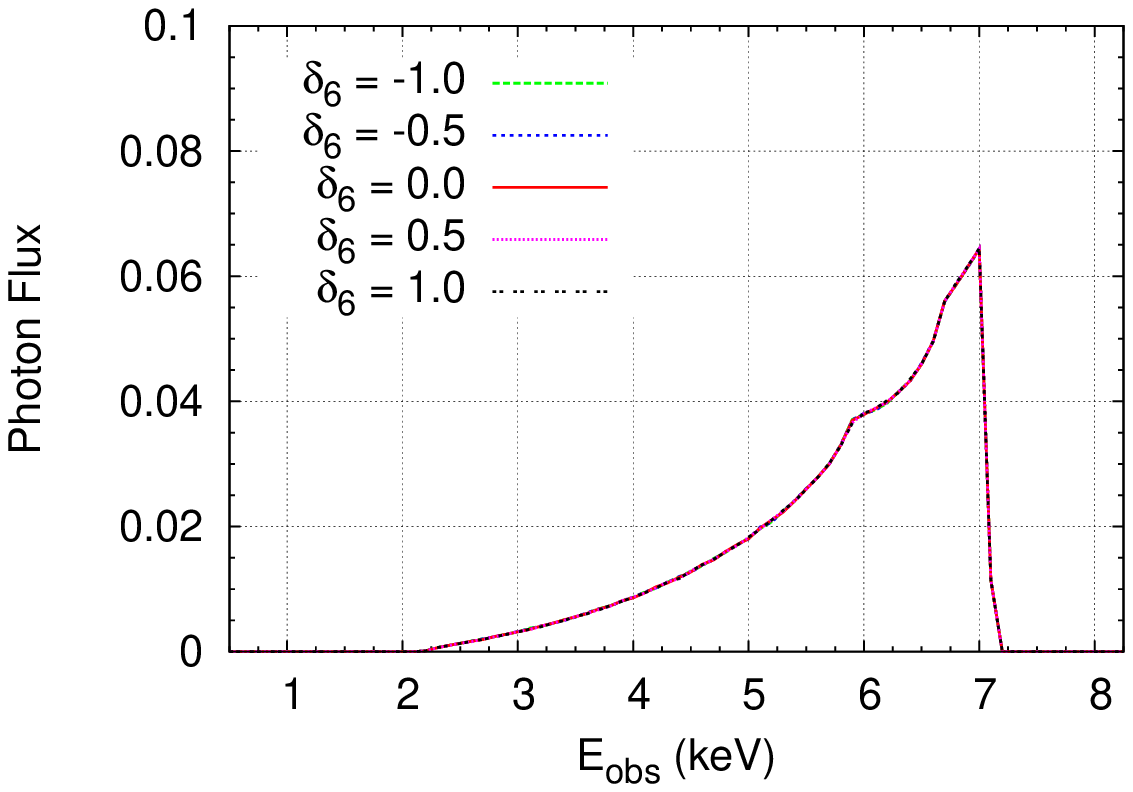}
\end{center}
\vspace{-0.3cm}
\caption{Impact of KRZ deformation parameters on the profile of the iron line. In each panel, one of the deformation parameters is allowed to vary while the others are set to zero. In all plots, the spin parameter is $a_* = 0.7$ and the inclination angle is $i = 45^\circ$. \label{f-lines}}
\end{figure}

The iron lines generated by our code for the KRZ metric are shown in Fig.~\ref{f-lines}. In these plots, we have considered the spin parameter $a_* = 0.7$ and inclination angle $i = 45^\circ$. In each plot, one of the KRZ deformation parameters is allowed to vary while the other five are set to zero. The red solid line representing the Kerr case is the same in all panels. From this panels, it is clear that $\delta_1$ and $\delta_2$ have a strong impact on the iron line, $\delta_3$, $\delta_4$, and $\delta_5$ have a moderate effect, while $\delta_6$ has no or very weak impact on the shape of the iron line.

\section{Simulations \label{s-sim}}

The aim of this section is to study how present and future observations can constrain the values of the KRZ deformation parameters and thus test the Kerr black hole hypothesis. For this purpose, we do not consider a specific source, but we simply employ the typical parameters for a bright black hole binary, which is the best source for this kind of measurements. As an explorative study, we adopt a simple model with a single iron K$\alpha$ line added to the power-law continuum. The power-law continuum, representing the spectrum from the hot corona, is generated with the photon index $\Gamma = 2$. The iron line used in the simulations is generated by our code assuming the Kerr metric ($\delta_j = 0$ for all $j$) with the spin parameter $a_* = 0.7$ and the inclination angle $i = 45^\circ$. The emissivity index is 3, namely the intensity profile scales as $1/r^3_{\rm e}$. The energy flux of the source is $6 \cdot 10^{-9}$~ergs/s/cm$^2$ in the range 1-9~keV, and the equivalent width of the iron line is 230~eV.

As an example of a current X-ray mission, we consider NuSTAR. Its effective area at 6~keV is about 800~cm$^2$ and, more importantly, its instrument does not have the problem of pile-up, so it is particularly suitable for the study of bright black hole binaries. The energy resolution at 6~keV is about 400~eV. For the next generation of X-ray observatories, we consider eXTP, which is a China-Europe project currently scheduled to be launched in 2022. Among the expected four instruments on board of eXTP, we only consider LAD, which is the most suitable for the study of bright sources. It has an unprecedented effective area of more than 30,000~cm$^2$ at 6~keV~\cite{spie16}, which is clearly a significant advantage with respect to current X-ray missions for testing the Kerr metric. The energy resolution at 6~keV is expected to be better than 200~eV. For both NuSTAR and LAD/eXTP, we simulate observations assuming the exposure time $\tau = 100$~ks. The photon count in the 1-9~keV range turns out to be $N_{\rm ph} \sim 10^7$ for NuSTAR, and $N_{\rm ph} \sim 10^9$ for LAD/eXTP. This will clearly make a difference in the final result.

\begin{figure}[t]
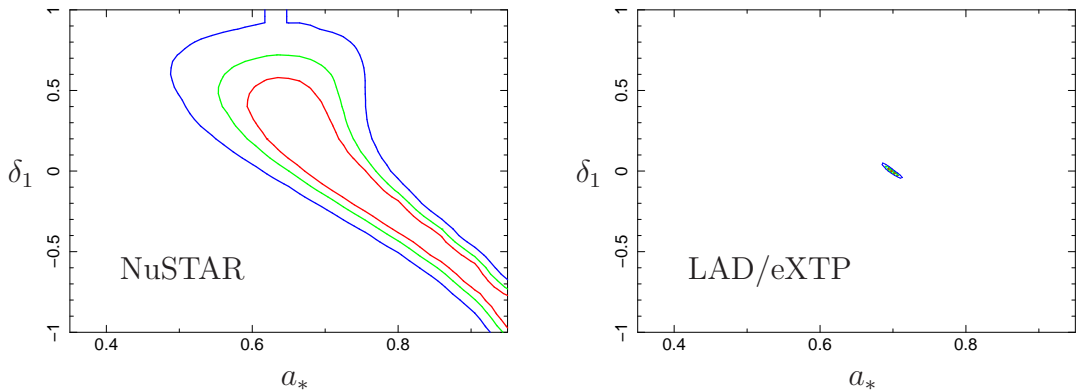

\vspace{0.4cm}
\begin{center}
\begin{overpic}[scale=0.27,angle=270]{d1_nustar_100ks.ps}
\put(-3,35){\large{$\delta_1$}}
\put(53,-7){\large{$a_*$}}
\put(20,15){\large{NuSTAR}}
\end{overpic}
\hspace{0.8cm}
\begin{overpic}[scale=0.27,angle=270]{d1_lad_100ks.ps}
\put(-3,35){\large{$\delta_1$}}
\put(53,-7){\large{$a_*$}}
\put(20,15){\large{LAD/eXTP}}
\end{overpic}
\end{center}
\vspace{0.4cm}
\caption{$\Delta\chi^2$ contours from simulations with NuSTAR (left panel) and LAD/eXTP (right panel) assuming an exposure time of 100~ks. The reference model is a Kerr black hole in which the spin parameter is $a_* = 0.7$ and the inclination angle is $i = 45^\circ$. It is compared with the predictions for spacetimes with non-vanishing deformation parameter $\delta_1$. The red, green, and blue curves indicate, respectively, the 1-, 2-, and 3-$\sigma$ confidence level limits. See the text for more details. \label{f-d1}}
\end{figure}

The simulated observations are treated as real data and fitted with the model 
\be
\verb|powerlaw + KRZ| \, , 
\ee
where \verb|KRZ| is a table model made by ourselves using XSPEC. It gives the iron line in the KRZ metric calculated by our own code, considering different spins, inclination angles, and deformation parameters. As a preliminary study, we only consider one non-vanishing deformation parameter at each time and set the others to zero. In our fits, we have six free parameters: the photon index of the power-law continuum $\Gamma$, the normalization of the continuum, the spin $a_*$, the inclination angle $i$, one of the six deformation parameters $\delta_j$, and the normalization of the iron line. Our results are summarized in Figs.~\ref{f-d1}-\ref{f-d6}, which show the contour levels of $\chi^2$ in the plane spin parameter vs deformation parameter.

\begin{figure}[t]
\vspace{0.4cm}
\begin{center}
\begin{overpic}[scale=0.27,angle=270]{d2_nustar_100ks.ps}
\put(-3,35){\large{$\delta_2$}}
\put(53,-7){\large{$a_*$}}
\put(20,15){\large{NuSTAR}}
\end{overpic}
\hspace{0.8cm}
\begin{overpic}[scale=0.27,angle=270]{d2_lad_100ks.ps}
\put(-3,35){\large{$\delta_2$}}
\put(53,-7){\large{$a_*$}}
\put(20,15){\large{LAD/eXTP}}
\end{overpic}
\end{center}
\vspace{0.4cm}
\caption{As in Fig.~\ref{f-d1} for the deformation parameter $\delta_2$. The red, green, and blue curves indicate, respectively, the 1-, 2-, and 3-$\sigma$ confidence level limits. See the text for more details. \label{f-d2}}
\vspace{1.0cm}
\begin{center}
\begin{overpic}[scale=0.27,angle=270]{d3_nustar_100ks.ps}
\put(-3,35){\large{$\delta_3$}}
\put(53,-7){\large{$a_*$}}
\put(15,15){\large{NuSTAR}}
\end{overpic}
\hspace{0.8cm}
\begin{overpic}[scale=0.27,angle=270]{d3_lad_100ks.ps}
\put(-3,35){\large{$\delta_3$}}
\put(53,-7){\large{$a_*$}}
\put(20,15){\large{LAD/eXTP}}
\end{overpic}
\end{center}
\vspace{0.4cm}
\caption{As in Fig.~\ref{f-d1} for the deformation parameter $\delta_3$. The red, green, and blue curves indicate, respectively, the 1-, 2-, and 3-$\sigma$ confidence level limits. See the text for more details. \label{f-d3}}
\end{figure}

\begin{figure}[t]
\vspace{0.4cm}
\begin{center}
\begin{overpic}[scale=0.27,angle=270]{d4_nustar_100ks.ps}
\put(-3,35){\large{$\delta_4$}}
\put(53,-7){\large{$a_*$}}
\put(15,15){\large{NuSTAR}}
\end{overpic}
\hspace{0.8cm}
\begin{overpic}[scale=0.27,angle=270]{d4_lad_100ks.ps}
\put(-3,35){\large{$\delta_4$}}
\put(53,-7){\large{$a_*$}}
\put(15,15){\large{LAD/eXTP}}
\end{overpic}
\end{center}
\vspace{0.4cm}
\caption{As in Fig.~\ref{f-d1} for the deformation parameter $\delta_4$. The red, green, and blue curves indicate, respectively, the 1-, 2-, and 3-$\sigma$ confidence level limits. See the text for more details. \label{f-d4}}
\vspace{1.0cm}
\begin{center}
\begin{overpic}[scale=0.27,angle=270]{d5_nustar_100ks.ps}
\put(-3,35){\large{$\delta_5$}}
\put(53,-7){\large{$a_*$}}
\put(15,25){\large{NuSTAR}}
\end{overpic}
\hspace{0.8cm}
\begin{overpic}[scale=0.27,angle=270]{d5_lad_100ks.ps}
\put(-3,35){\large{$\delta_5$}}
\put(53,-7){\large{$a_*$}}
\put(20,15){\large{LAD/eXTP}}
\end{overpic}
\end{center}
\vspace{0.4cm}
\caption{As in Fig.~\ref{f-d1} for the deformation parameter $\delta_5$. The red, green, and blue curves indicate, respectively, the 1-, 2-, and 3-$\sigma$ confidence level limits. See the text for more details. \label{f-d5}}
\end{figure}

\begin{figure}[t]
\vspace{0.4cm}
\begin{center}
\begin{overpic}[scale=0.27,angle=270]{d6_nustar_100ks.ps}
\put(-3,35){\large{$\delta_6$}}
\put(53,-7){\large{$a_*$}}
\put(47,55){\large{NuSTAR}}
\end{overpic}
\hspace{0.8cm}
\begin{overpic}[scale=0.27,angle=270]{d6_lad_100ks.ps}
\put(-3,35){\large{$\delta_6$}}
\put(53,-7){\large{$a_*$}}
\put(20,15){\large{LAD/eXTP}}
\end{overpic}
\end{center}
\vspace{0.4cm}
\caption{As in Fig.~\ref{f-d1} for the deformation parameter $\delta_6$. The red, green, and blue curves indicate, respectively, the 1-, 2-, and 3-$\sigma$ confidence level limits. See the text for more details. \label{f-d6}}
\end{figure}

The constraints on the deformation parameter $\delta_1$ are shown in Fig.~\ref{f-d1}, where the left panel refers to those with NuSTAR and the right panel to the constraints with LAD/eXTP. In the case of the NuSTAR observation, there is clearly a quite pronounced correlation between the estimate of the values of the spin and of the deformation parameter. The much higher photon count number in the LAD/eXTP measurement permits to break such a degeneracy and the constraints on both parameters look impressive.

Fig.~\ref{f-d2} shows the contour levels of $\chi^2$ for the deformation parameter $\delta_2$. Here we have a very strong correlation between the estimate of the spin and possible deviations from the Kerr solution. In the case of NUSTAR, the spin is almost unconstrained if we permit non-vanishing values of $\delta_2$. In the case of LAD/eXTP, the extended valley in the contour levels of $\chi^2$ shrinks to several small islands corresponding to several local minima of $\chi^2$.

The constraints on the deformation parameters $\delta_3$, $\delta_4$, and $\delta_5$ are illustrated in Figs.~\ref{f-d3}-\ref{f-d5}. These three parameters mainly influence the profile of the iron line, while they do not alter very much the low energy tail of the line. For this reason, the correlation between the estimate of the spin and of the deformation parameter is only moderate. With NuSTAR, it is impossible to get a constraint on these deformation parameters, which can be either larger than 1 or smaller than $-1$. In the case of LAD/eXTP, thanks to its superb effective area, all these parameters can be constrained quite well, with very stringent bound for $\delta_5$ and weaker bounds for $\delta_3$.

The constraints on the deformation parameter $\delta_6$ are shown in Fig.~\ref{f-d6}. Such a deformation parameter is very elusive, as it was already clear from Fig.~\ref{f-lines}. It does not appreciably affect the shape of the iron line, and even in the case of LAD/eXTP we do not get any meaningful bound on its value. It would be interesting to see whether such a deformation parameter could be constrained by other techniques, which are sensitive to different relativistic effects.

\section{Summary and conclusions \label{s-last}}

The iron K$\alpha$ line in the X-ray reflection spectrum of astrophysical black holes is generated from the inner part of the accretion disk and may be a powerful tool to probe the strong gravity region. In this paper, we have extended previous work in order to use this technique to test the Kerr black hole hypothesis. We have employed the parametrization recently proposed by Konoplya, Rezzolla, and Zhidenko in Ref.~\cite{krz}, which has some subtle advantages with respect to the existing parametrizations commonly used in this research field. We have simulated observations for a generic black hole binary, which is a more suitable source than an AGN for testing the Kerr metric. We have considered two sets of simulations: one to illustrate the capability of current X-ray missions, the other one to figure out that offered by the next generation of X-ray observatories. For the first set of observations, we have chosen NuSTAR, which has a larger effective area than other current X-ray missions at 6~keV and does not have the problem of pile-up in the case of bright black hole binaries. As a future observational facility, we have considered eXTP, which is a China-Europe project and is currently scheduled to be launched in 2022.

Our results are shown in Figs.~\ref{f-d1}-\ref{f-d6}. The constraining power between NuSTAR and eXTP is impressive. In the case of NuSTAR, it is impossible to constrain any deformation parameter of the KRZ metric. Large deviations for the Kerr solution cannot be ruled out. In the case of eXTP, for the same source and the same exposure time we find very stringent constraints on almost every deformation parameter. The exception is $\delta_6$, which does not seem to produce any appreciable effect on the iron line shape. It would be interesting to see if other observations, sensitive to different relativistic effects, can constrain $\delta_6$ or if such a deformation parameter remains elusive and it is difficult to measure. For the other five deformation parameters of the KRZ metric, we find that it is possible to constrain the deformation parameter at the level of 0.1 with the parameters chosen in our simulations (energy flux around $10^{-9}$~erg/s/cm$^2$ in the 1-9~keV range, iron line equivalent width around 200~eV, and exposure time of 100~ks).

The difference between NuSTAR and eXTP is mainly due to the very large effective area of the LAD instrument. The effective area of NuSTAR at 6~keV is about 800~cm$^2$, while LAD has an effective area at 6~keV of more than 30,000~cm$^2$. Actually the constraints found with LAD/eXTP are so stringent that systematics effects may be dominant. For this purpose, it will be very important to fit the future X-ray data with sophisticated theoretical models. Current uncertainties, in particular concerning the behavior of the emissivity profile, may prevent or limit the possibility of performing accurate tests of the Kerr metric.

In our analysis, we have considered several simplifications, which should be removed in future work and when we will analyze real data. In particular, the X-ray spectrum has been modeled by a simple power law and an iron line. While the iron line is the most informative feature for the metric in the strong gravity region, one has to fit the whole reflection spectrum. The version of the code used in this work is also not suitable to scan a model with several free parameters, and this strongly limits the ability to fit real data. In a future paper, we plan to adopt the strategy discussed in Ref.~\cite{full} and eventually obtain the extension of the relativistic reflection model RELXILL with the KRZ proposal as background metric.

%%%%%%%%%%%%%%%%%%%%%%%%%%%%%%%

\begin{acknowledgments}
This work was supported by the NSFC (grants 11305038 and U1531117) and the Thousand Young Talents Program. C.B. also acknowledges support from the Alexander von Humboldt Foundation.
\end{acknowledgments}

%%%%%%%%%%%%%%%%%%%%%%%%%%%%%%%

\end{document}